\documentclass[10pt,letterpaper]{article}
\usepackage[top=0.85in,left=2.75in,footskip=0.75in]{geometry}
\usepackage{upgreek}
\usepackage{amsmath,amssymb}
\usepackage{graphicx}
\graphicspath{{figresubmit/}}
\usepackage{changepage}

\usepackage[utf8x]{inputenc}

\usepackage{textcomp,marvosym}

\usepackage{cite}

\usepackage{nameref,hyperref}

\usepackage[right]{lineno}

\usepackage{microtype}
\DisableLigatures[f]{encoding = *, family = * }

\usepackage[table]{xcolor}

\usepackage{array}

\newcolumntype{+}{!{\vrule width 2pt}}
\newcommand{\fig}{Fig }

\newlength\savedwidth

\raggedright
\usepackage[aboveskip=1pt,labelfont=bf,labelsep=period,justification=raggedright,singlelinecheck=off]{caption}

\makeatletter
\renewcommand{\@biblabel}[1]{\quad#1.}
\makeatother

\date{}

\begin{document}
\vspace*{0.2in}

\begin{flushleft}
{\Large
\textbf\newline{DNA-Graphene Interactions During Translocation Through Nanogaps} % Please use "title case" (capitalize all terms in the title except conjunctions, prepositions, and articles).
}
\newline
\\
Hiral N.~Patel\textsuperscript{1},
Ian Carroll\textsuperscript{1},
Rodolfo Lopez, Jr.\textsuperscript{1},
Sandeep Sankararaman\textsuperscript{1},
Charles Etienne\textsuperscript{1},
Subba Ramaiah Kodigala\textsuperscript{1},
Mark R.~Paul\textsuperscript{2},
Henk W.Ch.~Postma\textsuperscript{1,*}
\\
\bigskip
\textbf{1} Department of Physics and Astronomy, California State University Northridge, Northridge, California, United States of America
\\
\textbf{2} Department of Mechanical Engineering, Virginia Polytechnic Institute and State University, Blacksburg, Virginia, United States of America
\\
\bigskip

% Insert additional author notes using the symbols described below. Insert symbol callouts after author names as necessary.
% 
% Remove or comment out the author notes below if they aren't used.
%
% Primary Equal Contribution Note
%\Yinyang These authors contributed equally to this work.
%
%% Additional Equal Contribution Note
%% Also use this double-dagger symbol for special authorship notes, such as senior authorship.
%\ddag These authors also contributed equally to this work.
%
%% Current address notes
%\textcurrency Current Address: Dept/Program/Center, Institution Name, City, State, Country % change symbol to "\textcurrency a" if more than one current address note
%% \textcurrency b Insert second current address 
%% \textcurrency c Insert third current address
%
%% Deceased author note
%% Group/Consortium Author Note
%\textpilcrow Membership list can be found in the Acknowledgments section.

% Use the asterisk to denote corresponding authorship and provide email address in note below.
* postma@csun.edu

\vspace{0.5cm}
\underline{This manuscript is published in PLOS One, see  doi.org/10.1371/journal.pone.0171505}

\end{flushleft}
\section*{Abstract}
We study how double-stranded DNA translocates through graphene nanogaps. Nanogaps are fabricated with a novel capillary-force induced graphene nanogap formation technique. DNA translocation signatures for nanogaps are qualitatively different from those obtained with circular nanopores, owing to the distinct shape of the gaps discussed here. Translocation time and conductance values vary by $\sim 100$\%, which we suggest are caused by local gap width variations. We also observe exponentially relaxing current traces. We suggest that slow relaxation of the graphene membrane following DNA translocation may be responsible. We conclude that DNA-graphene interactions are important, and need to be considered for graphene-nanogap based devices. This work further opens up new avenues for direct read of single molecule activitities, and possibly sequencing. 

%\linenumbers

% Use "Eq" instead of "Equation" for equation citations.
\section*{Introduction}
Solid-state and biological nanopores hold great promise as analytical single-molecule tools \cite{muthukumar_single-molecule_2015}. They enable study of folding dynamics \cite{bundschuh_coupled_2005}, enzyme activity \cite{fennouri_single_2012}, direct detection of DNA knots \cite{plesa_direct_2016}, and detection of single-nucleotide polymorphisms \cite{zhao_detecting_2007}. They may even enable direct-read single-molecule sequencing \cite{branton_potential_2008}. Solid-state nanopores may be fabricated with a focused ion beam \cite{li_ion-beam_2001}, atomic force microscope \cite{held_nanolithography_1998}, transmission electron microscope \cite{storm_fabrication_2003}, or dielectric breakdown \cite{kwok_nanopore_2014,yanagi_fabricating_2014,kuan_electrical_2015}. Graphene has especially advantageous properties as a material for nanopore studies  \cite{schneider_dna_2010,merchant_dna_2010,garaj_graphene_2010}.  First results for the MiniIon nanopore sequencer are promising, but show a relatively high error rate \cite{jain_improved_2015}. While this may be improved by repeated sequencing of identical molecules, this means there is still an unmet need in single-molecule {\em de novo} sequencing. Graphene nanogaps are a promising candidate for such a 
sequencing device \cite{postma_rapid_2010,he_enhanced_2011,prasongkit_transverse_2011, heerema_graphene_2016}. 

Here, we present the first studies of double-stranded DNA (dsDNA) translocating through graphene nanogaps. The translocation signatures differ significantly from those found in other solid-state and biological nanopores, owing to the unique properties of these graphene nanogaps, and unique DNA-graphene nanogap interactions. Our nanogap formation procedure is based on capillary-force-induced breaking  that can be controlled down to a few nanometers.

\section*{Methods}

\begin{figure}[h]
\includegraphics[scale=0.5]{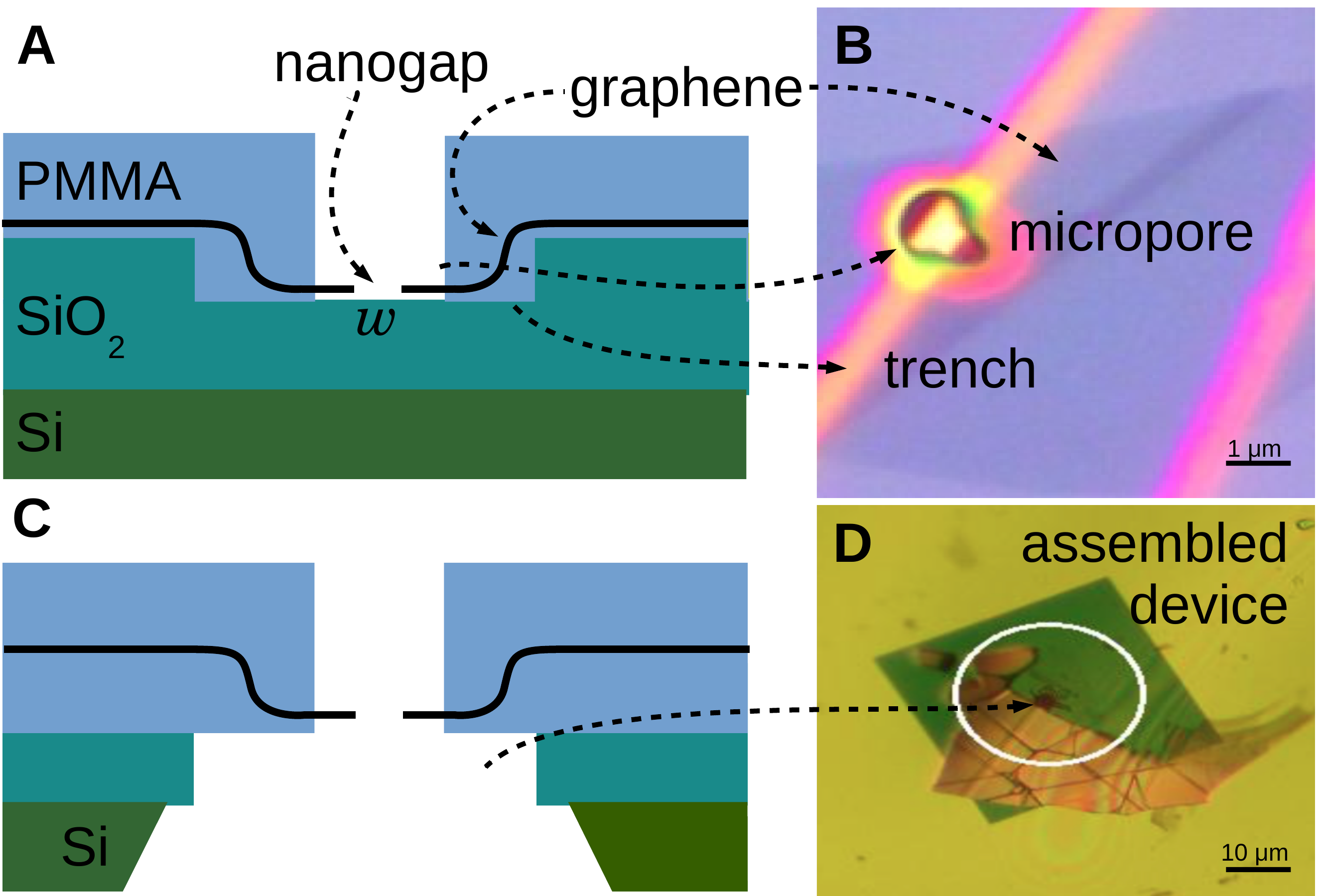}
\caption{\label{fig1} {\bf Fabrication procedure for the graphene nanogap devices studied here.} (A) Diagram of graphene placed by wedge transfer on top of prefabricated trench in SiO\textsubscript{2} substrate. A nanogap with width $w$ is formed. A micropore is fabricated by electron-beam lithography over the nanogap location. (B) Optical microscope image of a graphene nanogap device with micropore at the same fabrication stage as \fig \ref{fig1}A, placed on an etched trench (yellow) in a $300$ nm thick SiO\textsubscript{2} substrate (purple). Contrast has been enhanced. The fabricated micropore changes the color of the SiO\textsubscript{2} substrate and trench, making the trench appear brighter yellow and the fraction of uncovered unetched SiO\textsubscript{2} more purple. The image has 50\% increased contrast and 30\% reduced brightness. (C) Diagram of a graphene nanogap transferred to a  measurement device with a $\sim 25 \; \upmu \mathrm{m}$ hole that the nanogap/micropore assembly is centered on. (D)  Optical microscope image with enhanced contrast of a graphene nanogap device ready for transverse conductance measurements at the fabrication stage shown in \fig \ref{fig1}C. }
\end{figure}

We fabricate graphene nanogaps and demonstrate that DNA is able to translocate through them (\fig \ref{fig1}).

Graphene is deposited on Si/SiO\textsubscript{2} wafers as described before \cite{schneider_wedging_2010}. A \mbox{$5-10$ mm} piece of a graphenium flake (manufacturer: NGS Naturgraphit GmbH) is mechanically exfoliated with Blue Nitto tape (manufacturer: Nitto Denko, SPV 224LB-PE). The flakes are deposited on a Si wafer with a thermally-grown $285$ nm thick SiO\textsubscript{2} layer that has been treated with a $5$ min long O\textsubscript{2} plasma to promote graphene adhesion. A single-layer graphene sheet is identified by its contrast in an optical microscope. 

Si/SiO\textsubscript{2} wafers with trenches and buried metal alignment markers are fabricated using  electron-beam lithography (EBL), processing, and etching. Briefly, PMMA is patterned by writing with an electron beam and subsequent development in a 1:3 mixture of Methyl Isobutyl Ketone (MIBK) and Isopropyl Alcohol (IPA). We thereby created a \mbox{$20\times 20 \; \upmu \mathrm{m}^2$}  square mask, which we etch for $30$ s in buffered oxide etchant (BHF, manufacturer: J.T. Baker). A $25$ nm thin layer of Au is evaporated into the squares after a \mbox{$3$ nm} Cr adhesion layer, followed by liftoff in \mbox{$80 \; ^\circ$C} acetone. A second layer of PMMA resist is applied, followed by EBL-definition of $1\times 100 \; \upmu \mathrm{m}^2$ trenches and development. The trenches are made \mbox{$\sim 150$ nm} deep by subsequent etching in BHF. 

Exfoliated single-layer graphene sheets are transferred onto the trenches using a wedge-transfer technique (\fig \ref{fig1}A,B) developed by Schneider {\em et al.}\cite{schneider_wedging_2010}. Graphene sheets are covered in cellulose acetate butyrate (CAB) in ethyl acetate (EtAc)  and wedged by water. They are positioned over the trench, after which the water level is lowered. The CAB membrane is dissolved in EtAc, and the solvent is exchanged for IPA, after which the sample is dried with N\textsubscript{2} gas. The capillary forces of the drying IPA break the graphene sheet inside of the trench, thus forming a graphene nanogap \cite{wang_fabrication_2010}. The size of the gap is determined by the balance between the bending energy of the graphene sheet and its adhesion to the underlying SiO\textsubscript{2} substrate, and can be controlled down to a few nanometers, as we describe below.

To create a narrow channel to guide the DNA to the nanogap, and to block any other potential holes that may have been accidentally created, a \mbox{$\sim \upmu \mathrm{m}$-size} pore in PMMA, ``micropore'', is fabricated on top of the graphene nanogap (\fig \ref{fig1}A,B). The coordinates of the graphene nanogap with respect to the previously defined Au square markers are determined in the optical microscope. A PMMA film is deposited, and the micropore is exposed over the graphene nanogap using custom-written EBL software and development (\fig \ref{fig1}A,B). The PMMA film containing the micropore over the graphene nanogap is then transferred to a prefabricated hole in a SiN membrane (\fig \ref{fig1}C,D). 

The assembly is mounted in a fluid cell with a saline solution of \mbox{$10$ mM} Tris acetate, \mbox{$1$ mM} Ethylenediaminetetraacetic acid (EDTA) and \mbox{$0.1$ M} KCl. Ag/AgCl electrodes are mounted on either side, and a $\sim 20\mbox{ mV}$ bias voltage is applied. The nanogaps used here can be expected to have a larger cross section and a corresponding larger conductance than the graphene nanopores reported earlier \cite{schneider_dna_2010,merchant_dna_2010,garaj_graphene_2010}. In order to avoid saturating our current amplifier, we used a lower salt concentration than the \mbox{$1-3$ M} in previously-reported graphene nanopore studies \cite{schneider_dna_2010,merchant_dna_2010,garaj_graphene_2010}.  The ion current is amplified with a patch-clamp amplifier (Axopatch 200B) and recorded at a sample rate of \mbox{$1.25$ MHz} with custom software \cite{postma_daq:_2012}.

Double-stranded DNA ($\uplambda$-DNA, \mbox{$48$ kbp} long, Promega) is introduced on the cis side, and upon translocation through the nanogap, causes abrupt changes in ion-current.
% Results and Discussion can be combined.

\subsection*{Nanogap Size Calculations}

\begin{figure}[h]
\includegraphics[scale=0.5]{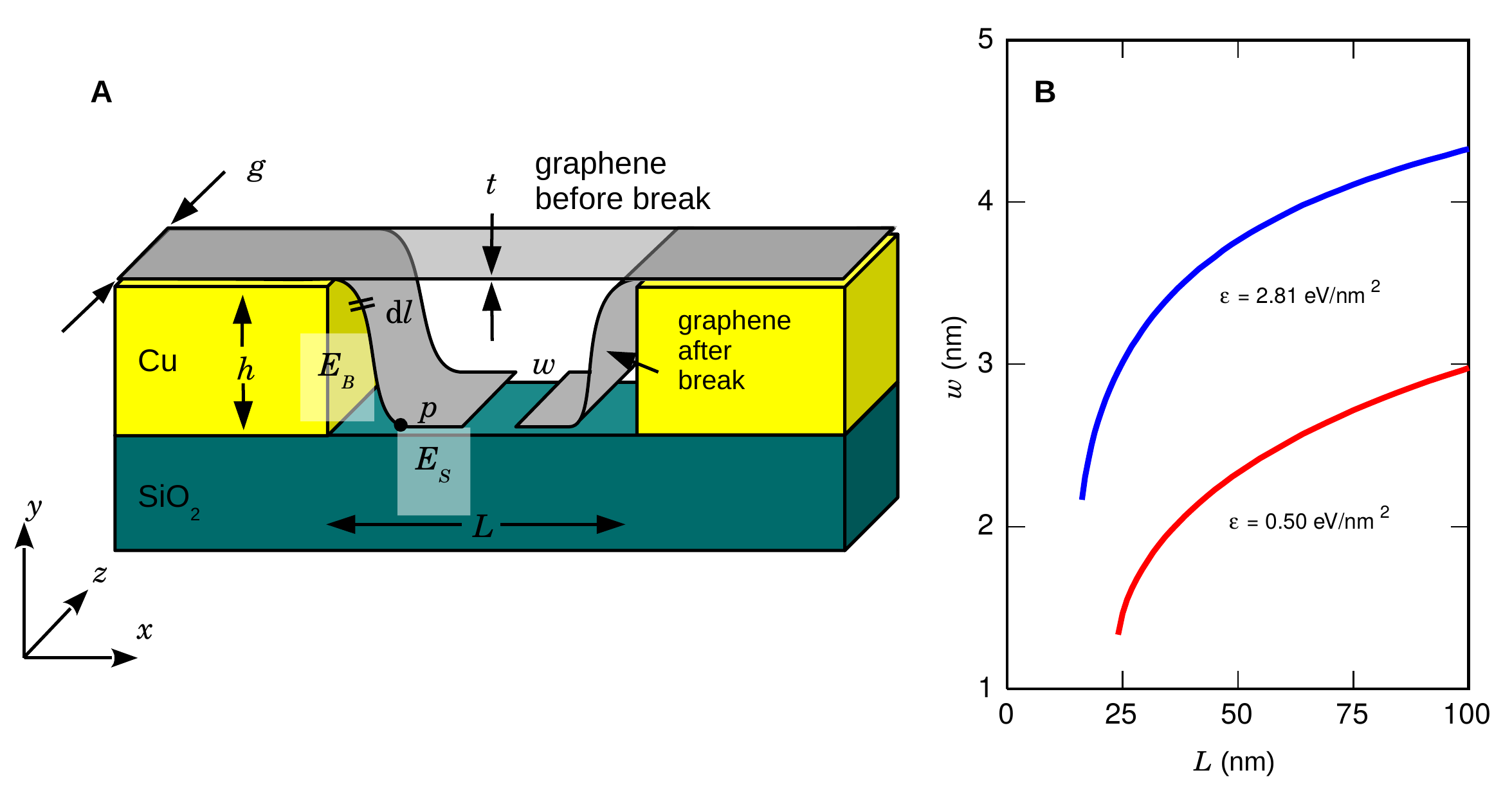}
\caption{\label{fig2} {\bf Model for breaking a graphene sheet}. (A) Breaking a graphene sheet of thickness $t$ over a trench of width $L$ and depth $h$. The graphene shape is estimated by balancing the bending energy $E_B$ up to the point $p$ with the surface binding energy $E_S$, leading to a nanogap width $w$. (B) Expected nanogap width $w$ as a function of suspended length $L$ for two values of surface adhesion energy $\varepsilon$ (see text). }
\end{figure}

We calculate the expected size of the nanogap by balancing the bending energy cost $E_B$ with a surface binding energy gain $E_S$ (\fig \ref{fig2}). Graphene adheres well to SiO\textsubscript{2} surfaces, but for a part of the graphene sheet to adhere to the underlying surface, the sheet has to be bent. The bending energy is stored in the bent portion of the graphene sheet between the support and the point of contact with the bottom of the trench,
\[
E_B = \frac{EI}{2}\int_x \left[ \left( \frac{\partial x}{\partial l} \right)^2 \frac{\partial ^2y}{\partial x^2} + \frac{\partial y}{\partial x} \frac{\partial ^2 x}{\partial l^2}\right]^2 \frac{\partial  x}{\partial l} \mathrm{d}x \qquad ,
\]
where \mbox{$E = 1$ TPa} is the Young's modulus \cite{bunch_electromechanical_2007}, $I = gt^3/12$ is the moment of inertia, \mbox{$t = 0.34$ nm} is graphene's thickness, and $g$ is the width of the graphene sheet. At equilibrium, the bending energy is equal to the surface binding energy $E_S = \varepsilon gs$, where $s$ is the length of the graphene sheet that is adhering to the SiO\textsubscript{2} substrate, and $\varepsilon = E_S/A$ is the surface adhesion energy. We model the nanogap size for both the SiO\textsubscript{2} adhesion energy of \mbox{$\varepsilon = 2.81$ eV/nm\textsuperscript{2}}  of graphene to SiO\textsubscript{2} \cite{koenig_ultrastrong_2011} and the graphene exfoliation energy $\varepsilon = 0.50 \; \mathrm{eV/nm}$\textsuperscript{2} \cite{zacharia_interlayer_2004}. We numerically solve these equations to deduce the nanogap width as a function of suspended length $L$ (\mbox{\fig \ref{fig2}B}).

In order to break the graphene sheet, we need to exceed the yield strength of graphene. In the experiments presented here, we broke the sheet in a trench with \mbox{$h = 150$ nm} and \mbox{$L = 1 \; \upmu \mathrm{m}$}. We numerically deduce $\Delta L/L = 49\%$ in this geometry, well above the failure strain of graphene of $25\%$ \cite{wei_nonlinear_2009}. Upon wedge transfer to the final device for the translocation studies presented here, we release it from the SiO\textsubscript{2} surface. Therefore, the graphene sheet can be expected to relax and $w$ to become smaller.

\subsection*{Scanning Electron Microscope Nonlinear Correction}

\begin{figure}[h]
\includegraphics[scale=0.9]{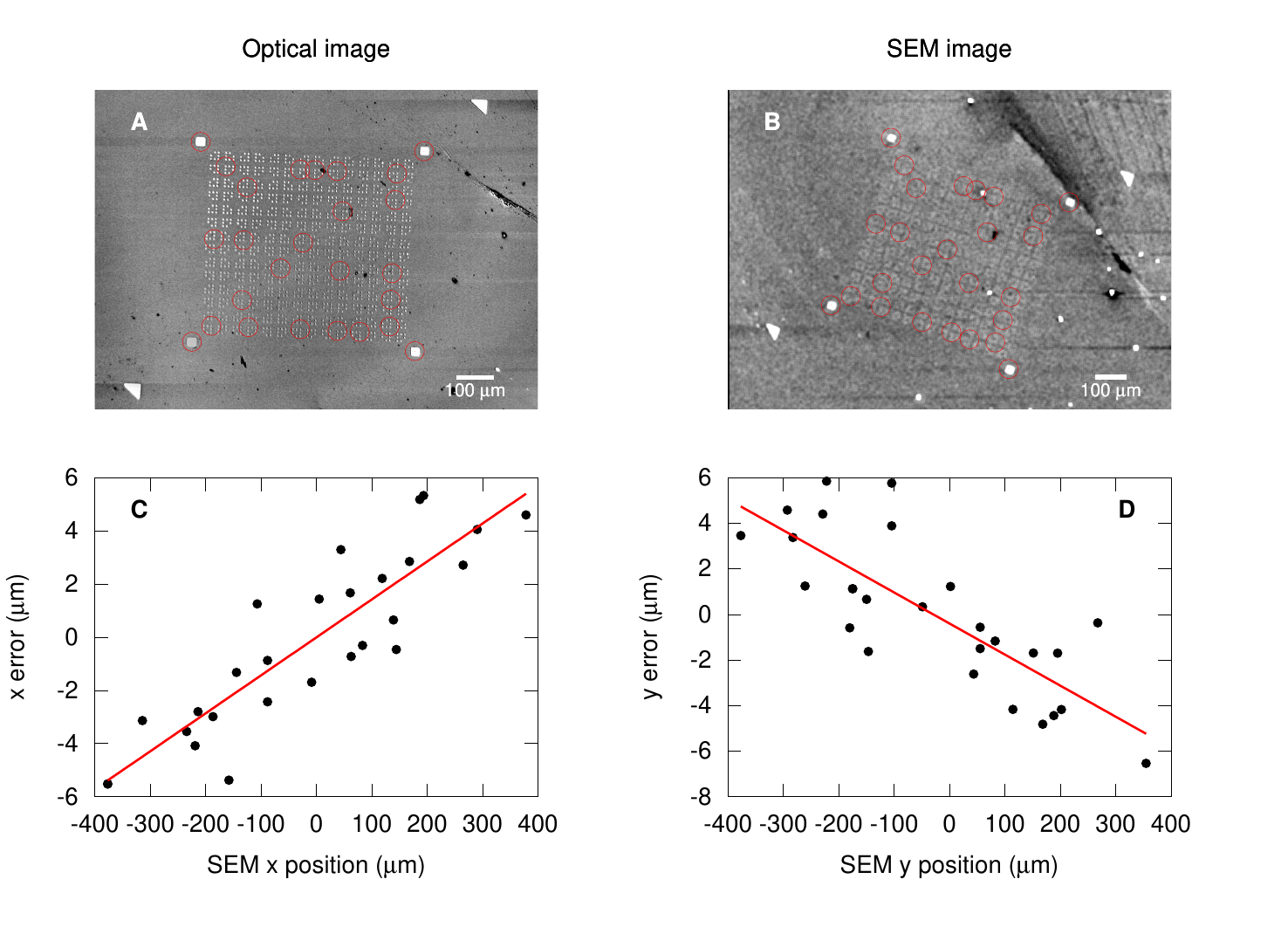}
\caption{\label{fig3} {\bf Nonlinear correction of electron-beam deflection.} (A) Optical image of Au alignment markers and selected markers with enhanced contrast for coordinate analysis (red circles). (B) SEM image of same device as A and selected markers with enhanced contrast  (red circles). (C) Residual error in $x$ coordinate of SEM beam deflection (black circles) that scales linearly with the SEM $x$ coordinate (red line). (D) Residual error (black circles) in SEM $y$ coordinate scaling linearly with the SEM $y$ coordinate (red line).
 } 
\end{figure}

The scanning electron microscope (JEOL JSM-840) is used to make a micropore in the PMMA above the graphene sheet. In order to locate the sheet with high accuracy, we calibrated the SEM beam deflection as a function of applied deflector voltage. A pattern of prefabricated Au square markers was imaged in an optical microscope (\mbox{\fig \ref{fig3}A}) and the SEM (\mbox{\fig \ref{fig3}B}). The $x$ and $y$ coordinates of markers in both optical and SEM images are recorded as complex numbers $z = x+\imath y$ (red circles). A linear fit of $z_{\mbox{\tiny SEM}}$ vs $z_{\mbox{\tiny OPT}}$ yields a best fit linear transform $z_{\mbox{\tiny SEM}} = a z_{\mbox{\tiny OPT}} + b$. This fit simultaneously records offset $b$, scale $|a|$, and rotation $\arg(a)$. The error after the fit $z_{\mbox{\tiny SEM}} - a z_{\mbox{\tiny OPT}} - b$ is analyzed \mbox{(\fig \ref{fig3}C,D)}. A small error remains that is linearly proportional to the $x$ and $y$ SEM coordinates. The correction is quadrupolar in nature; it is positive in the $x$ direction, and negative in the $y$ direction. We assume the most significant contribution to it is due to the electron optics in the SEM. The error can amount to several $\upmu \mathrm{m}$, and neglecting it would cause the micropore to not be located exactly over the trench. This quadrupolar correction was applied to our SEM micropore fabrication procedure. 

\subsection*{Control Experiments, Data Processing, and Event Detection}

\begin{figure}[h]
\includegraphics[scale=0.9]{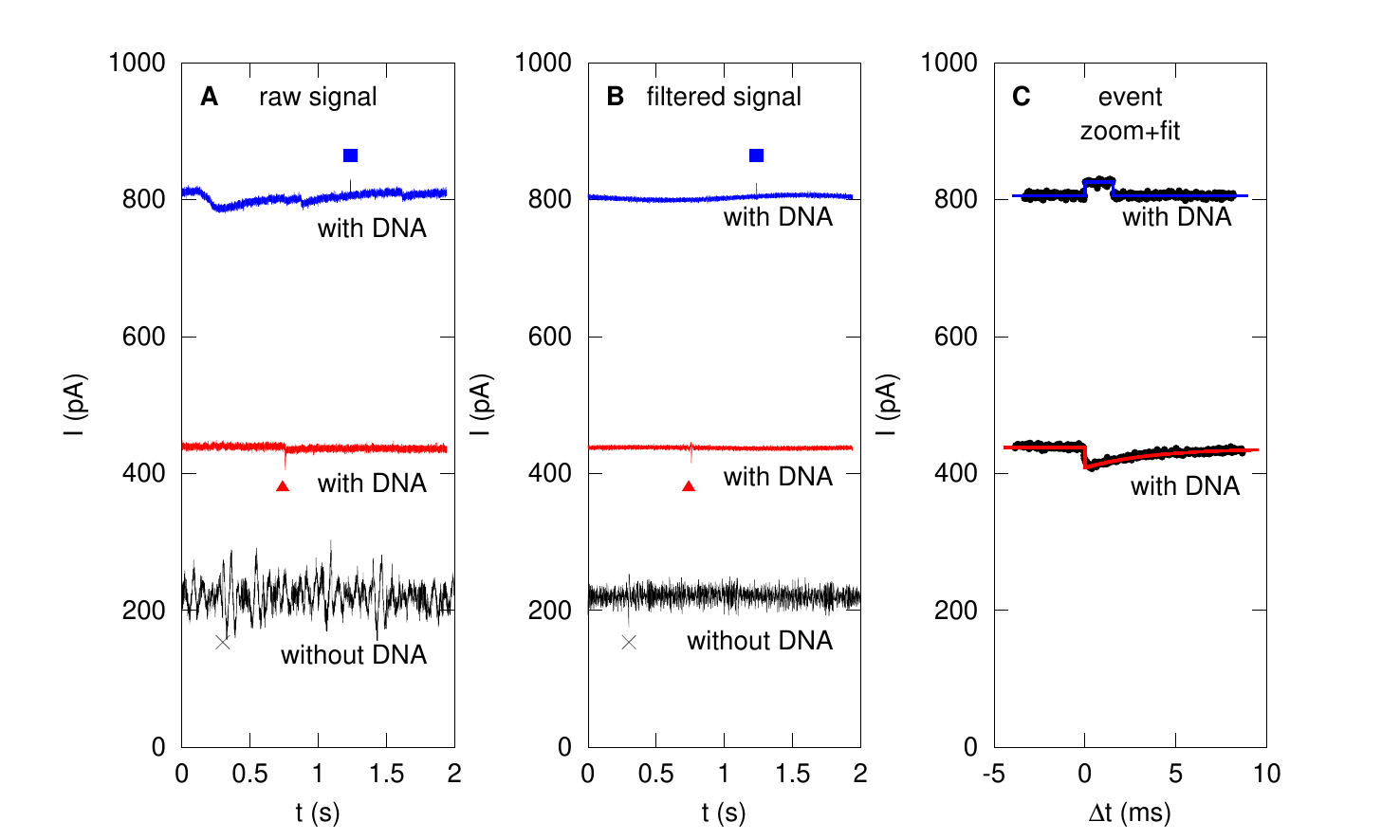}
\caption{\label{fig0} {\bf Event detection and classification}. (A) Raw current traces are recorded as described without DNA (black) and with DNA (red, blue). Eventual determination of rectangular and exponential events are indicated by blue square and red triangle, respectively. A dip in the control experiment ($\times$) is discarded as a possible event as described in the text. (B) Narrow-band noise is removed from the traces and eventual event position is indicated. (C) Events are fitted with rectangular (blue) or exponential (red) traces depending on which fits better statistically.}
\end{figure}

Control current traces are recorded before introduction of DNA, and traces with DNA are recorded as described above (\fig \ref{fig0}A). Translocation events are identified and analyzed in a three-step process as follows. First, the raw current traces are FFT filtered to remove spurious interference from instrumentation and environment (\fig \ref{fig0}B). Next, a slowly varying background is extracted from the signal by computing a running average using a one-second length. This background is subtracted from the signal, and the RMS value of the trace $\sigma$ is calculated. Candidate events are determined as either positive or negative peaks exceeding $5 \sigma$. The area under the peak is calculated to determine the statistical significance of the candidate event. The threshold for calling it an `event' is adjusted until the control experiments do not yield false positives anymore. We include particularly noisy control traces such as the black trace in \fig \ref{fig0}A to make sure that obviously-apparent intereference and spikes and dips do not yield any false positives either. Short dips such as that marked $\times$ in \fig \ref{fig0}A and \fig \ref{fig0}B are thereby excluded. Third, the {\em unfiltered} candidate events are fitted by a least squares method to both an exponentially decaying function and a rectangular function. We fit the unfiltered data to avoid distortions to the event shape by the FFT filtering and background subtraction. The squared difference between fit and data for both is used to determine whether the event better fits a rectangular or exponential event, and it is classified as such. The resulting fits are recorded and zoomed plots are generated (\fig \ref{fig0}C). All fits are visually confirmed. 

\section*{Results}

\begin{figure}[h]
\includegraphics[scale=1]{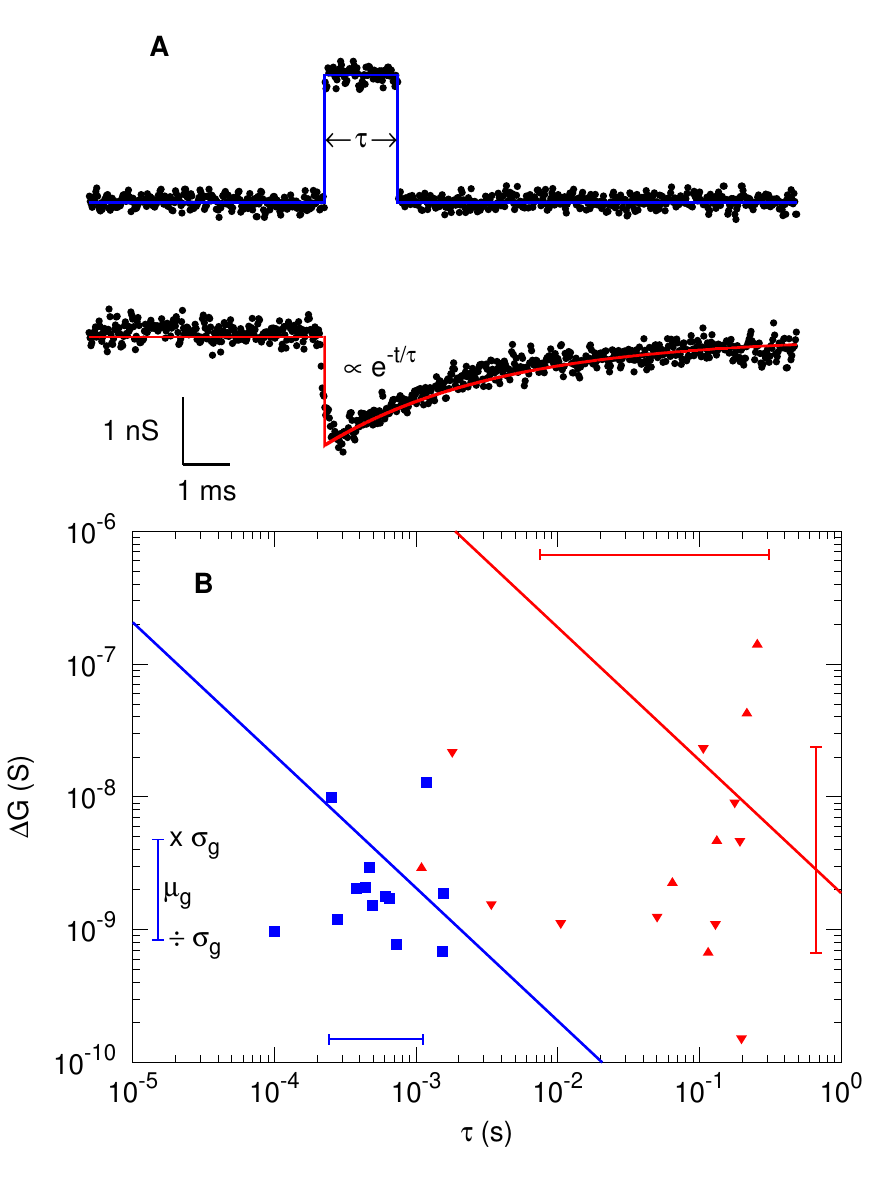}
\caption{\label{fig4} {\bf Recorded translocation of dsDNA through graphene nanogap with resistance of \mbox{$\sim 36 \; \mathrm{M}\Omega$}}. (A) Typical rectangular (blue, top) and exponential (red, bottom) translocation event. (B) Analysis of rectangular (blue squares) and exponential translocation events (red triangles, pointing up for $\Delta G > 0$ and down for $\Delta G < 0$). The solid lines are fits of the data to $\Delta G \propto 1/\tau$. The error bars indicate the range of geometrical standard deviation. }
\end{figure}

Upon introduction of dsDNA on the cis side of the chamber and application of a bias voltage between the two Ag/AgCl electrodes, brief changes in the ion current are observed (\fig \ref{fig4}A). These events only occur after introduction of the dsDNA, we therefore attribute them to translocation of dsDNA through the graphene nanogap. 

We observe two types of events. Firstly, we observe  rectangular events, where the current briefly changes from the baseline current (\fig \ref{fig4}A, top). The average change in conductance during such a blockade event, $\Delta G$, is $3$ nS, while the events have an average duration $\tau = 0.7$ ms (\fig \ref{fig4}B). These events have a geometric standard deviation $\sigma_g (x) = \exp(\mathrm{std}(\log(x)))$ in both $\Delta G$ and $\tau$ of $\sigma_g \sim 2$. For all rectangular events, we find that the conductance increases during the event, i.e. $\Delta G > 0$. 

Secondly, we observe short decay events, where the current rapidly changes from the baseline current and relaxes exponentially back with a decay constant $\tau$ ranging from $\sim 1$ to $100$ ms (\mbox{\fig \ref{fig4}A, bottom}). The maximum conductance change $\Delta G$ during such exponential events is larger than for rectangular events with an average value of $\Delta G = 20$ nS. These events have a geometric standard deviation in both $\Delta G$ and $\tau$ of $\sigma_g \sim 6$.  We find that during exponential events, the conductance change can be either positive or negative, with no discernible correlation between $\tau$ and the sign of $\Delta G$, nor between the magnitude of $\Delta G$ and the sign of $\Delta G$ (\fig \ref{fig4}B). For both types of events, the data is not following the behavior $\Delta G \propto 1/\tau$ that has been observed for circular graphene nanopores \cite{schneider_dna_2010,merchant_dna_2010,garaj_graphene_2010} (solid lines), as we find $\Delta G \propto \tau^\alpha$ with $\alpha = 0.04 \pm 0.34$ and $\alpha = 0.09 \pm 0.27$ for rectangular and exponential events, respectively.

\section*{Discussion}
The rectangular events have similar $\tau$ and $\Delta G$ average values to events commonly observed in solid-state nanopores \cite{li_ion-beam_2001,storm_fast_2005}. They are signatures of DNA translocation; the DNA temporarily occupies the nanopore and consequently changes the ion current through the pore. 

We attribute the positive nature of these translocation events ($\Delta G> 0$) to the low salt concentration that we employed in these experiments. In Si-based semiconducting nanopores, a crossover between $\Delta G> 0$ to $\Delta G< 0$ occurs at a salt concentration of $0.4 \; \mathrm{M}$ \cite{smeets_salt_2006}. At low salt concentrations, the conductance during an event is dominated by the counterion current along the DNA molecule, while at high concentrations, the conductance is dominated by blocking of the pore cross section by the DNA molecule. Our observation that $\Delta G> 0$ at a concentration of $0.1$ M is consistent with that picture and leads us to conclude that the conductance change is dominated by counterion current along the DNA molecule. 

\begin{figure}[h]
\includegraphics[scale=1.3]{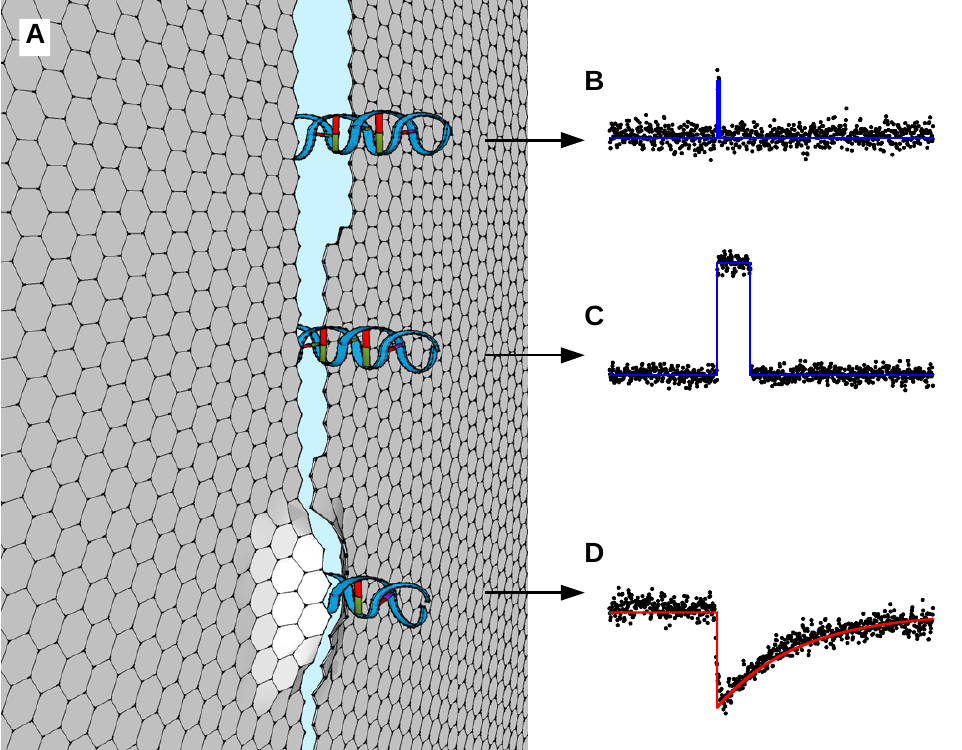}
\caption{\label{fig5} {\bf Effect of local nanogap geometry on translocation events, showing gaps and associated event data.} (A) Illustration of continuously changing gap width over the length of the gap. As DNA translocates through the gap, its passage is characterized by the gap width at its point of traversal. (B) DNA passes through the upper gap, causing a rectangular conductance change. (C) DNA passes through the middle gap more slowly due to the narrower gap diameter. (D) DNA forces the edges of the lower gap to bend outward, which then relax to their equilibrium position once it has passed, causing an exponential decay event.} 
\end{figure}

We attribute the large range in $\tau$ and $\Delta G$ to the unique properties of the rectangular graphene nanogaps studied here. In circular graphene nanopores, the ion current density profile of a graphene nanopore depends on the distance from the pore wall. The $\Delta G$ is therefore a function of which part of the pore's cross section is blocked by the DNA during translocation. However, in these pores, the point of translocation through the nanopore is approximately equal for all translocation events, and consequently $\Delta G$ does not vary much. The typical geometric standard deviation for graphene nanopores is $\sigma_g \sim 1.2$, or, equivalently, $\Delta G$ only varies by $ \sim 20$\% \cite{schneider_dna_2010,merchant_dna_2010}. In contrast, the graphene nanogaps studied here can be assumed to have edges that are serrated on a nanometer scale, causing the local width at the DNA's translocation point to vary (\fig \ref{fig5}A). The current density that is blocked by the DNA during translocation therefore varies more, and we find a larger $\sigma_g \sim 2$, a variation of $100$\%. It was recently reported that an increase of nanopore diameter by only a factor $2$ causes an increase in translocation rate by an order of magnitude due to decreasing nanopore-DNA interaction strength \cite{wanunu_dna_2008}. Due to the serrated nature of the nanogap edge, the effective width at the local point of translocation can be expected to parametrically alter the DNA-gap interaction strength, and thereby the translocation speed (\mbox{\fig \ref{fig5}B,C}). Since both $\Delta G$ and $\tau$ vary more than for circular graphene nanopores, the process for the events described here is fundamentally different, and it is not surprising that the trend $\Delta G \propto 1/\tau$ is not followed. Finally, the interaction between the graphene edge and the DNA molecule may also be affected by the nature of the edge, and harnessing this interaction may enhance performance of DNA sequencing devices built on this principle \cite{he_enhanced_2011,amorim_boosting_2016,he_functionalized_2008}.

The exponential decay events reported here have not been reported for circular nanopores, nor do we see them in our control experiments. They could therefore be due to the unique geometry of the nanogaps presented here. One possible mechanism responsible could be reversible mechanical changes to the graphene upon interaction with DNA. In such a case, the DNA may arrive at a part of the nanogap where the width is too small to go through. The membrane could be forced to flex, allowing passage of the DNA molecule, after which the membrane relaxes (\fig \ref{fig5}D). As the translocation event does not require a complete unfolding of the molecule, the translocation event itself may be much shorter than events that require unfolding of the molecule (rectangular events). Indeed, we observe a very short exponential increase of the current, followed by a longer exponential decay. We therefore establish an upper limit to the unfolded translocation time of $\sim 15 \; \upmu \mathrm{s}$, the temporal resolution of the experiment.

The exponential relaxation after the DNA has passed has an average geometric relaxation time of \mbox{$50$ ms}; it is surprisingly large. Micro and nanoscale elastic objects placed in a viscous fluid can often be modeled as a simple harmonic oscillator~\cite{paul_stochastic_2004,paul_stochastic_2006}. For nanoscale elastic objects in a viscous fluid the dynamics are essentially inertialess and can be described as an overdamped simple harmonic oscillator which yields long relaxation times.  To accurately model this mechanical relaxation, a detailed treatment of viscoelastic drag of counterions on the graphene sheet is required, and is beyond the scope of this paper.  

\subsection*{Modeling the Slow Exponential Decay Events}

We will show that the long relaxation times measured in our experiments are consistent with the dynamics that would occur for a graphene sheet returning to equilibrium after being initially displaced by the DNA passing through the gap.   We represent the motion of the fundamental mode of oscillation of the graphene membrane as a simple harmonic oscillator as 
\begin{equation}
 m_f \ddot{x}(t) + \gamma \dot{x}(t) + k x(t) = 0,
 \label{eq:sho}
\end{equation}
where $x$ is the displacement of the membrane, $m_f$ is the equivalent mass which includes the mass of the graphene membrane and the added mass of the fluid that is in motion ($m_f \gg m$ where $m$ is the mass of the graphene alone), $\gamma$ in the viscous damping acting on the graphene by water,  $k$ is the equivalent spring constant of the graphene, and $t$ is time.

For this system, it is straightforward to show that the Reynolds number of the fluid flow is very small $\text{Re} \ll 1$ which indicates the dominance of viscous effects over inertia. In light of this, we expect the graphene membrane to act as an overdamped oscillator after being displaced which has been demonstrated for nanoscale cantilevers in fluid~\cite{paul_stochastic_2004}.

If we assume that the initial displacement of the graphene membrane is $x_0 = F_0/k$ where $F_0$ is an applied force, the return to equilibrium of the graphene membrane can be expressed as~\cite{paul_stochastic_2006}
\begin{equation*}
x(t) = x_0 \left( \frac{\lambda_+}{\lambda_+-\lambda_-}e^{\lambda_- t} + \frac{\lambda_-}{\lambda_--\lambda_+}e^{\lambda_+ t} \right)
\label{eq:full}
\end{equation*}
where
\begin{equation*}
\lambda_{\pm} = \omega_f \left( -\frac{1}{2Q} \pm
\sqrt{\frac{1}{4Q^2}- 1}\right) \sim - \omega_f Q^{\pm 1}.
\label{eq:lambda}
\end{equation*}
In these expressions $Q = m_f \omega_f / \gamma$ is the quality factor and $\omega_f$ is the resonant frequency of the graphene membrane  when immersed in the fluid. For nanoscale oscillating objects in fluid $\omega_f \ll \omega_0$ where $\omega_0$ is the resonant frequency in a vacuum.

For small $Q$ we have $|\lambda_-| \gg |\lambda_+|$ where $\lambda_+ \sim - \omega_f Q$ and $\lambda_- \sim - \omega_f/Q$ and the large-$t$ response is dominated by $\lambda_+$. The solution can then be represented as $x(t) = x_0 e^{\lambda_+ t } = x_0 e^{- t/\tau}$ where $\tau$ is the relaxation time and is given by $\tau = (\omega_f Q)^{-1}$.

If we assume that the membrane dynamics are similar to that of a wide elastic cantilever in fluid we can make the following analytical predictions to suggest the order of magnitude of the expected response for the graphene membrane. Using the approach described in Paul {\em et al.}\cite{paul_stochastic_2006} we can predict the values of $Q$, $\omega_f$ (and therefore $\tau$) given only the values of frequency parameter $R_0$, mass loading parameter $T_0$, and the resonant frequency in vacuum $\omega_0$.

For the graphene membrane we will use a Young's modulus of $E = 1 \times 10^{12}$ N/m\textsuperscript{2} and a density of $\rho_g = 2000$ kg/m\textsuperscript{3}.  For water we will use a density of $\rho_f = 1000$ kg/m\textsuperscript{3} and a dynamic viscosity of $\mu_f = 1 \times 10^{-3}$ $\text{kg}/\text{m}\cdot \text{s}$. We next assume the membrane can be represented as a cantilever with a length $L \approx 1 \; \upmu$m, width $W \approx L/2$, and a thickness of \mbox{$H \approx 0.3$ nm.}  Using these values yields an equivalent spring constant of $k \approx 3.4~\upmu\text{N/m}$ and a resonant frequency in vacuum of $\omega_0 \approx 6.8 \times 10^{6} \text{rad/s}$.

These values yield a frequency parameter $R_0 = \rho_f \omega_0 W^2/(4 \mu_f) = 0.43$ and a mass loading parameter of 
$T_0 = \pi \rho_f W/(4 \rho H) = 655$.   Given these values for $R_0$, $T_0$, and $\omega_0$ the analytical expressions of Paul {\em et al.}\cite{paul_stochastic_2006}  yield $Q \approx 0.17$ and $\omega_f \approx 1694$ rad/s.  Using this with our expression for the decay time yields $\tau \approx 3$ ms. This is commensurate with the long relaxation times that are measured in the experiments.  We emphasize that these estimates are approximate and a more accurate analysis would require numerical simulations for the precise conditions of the experiment. In addition, the predicted value of the relaxation time from our approximate analysis is sensitive to the chosen geometry of the membrane.  For example, as the length or width of the membrane becomes larger the decay time will increase.

\section*{Conclusion}

We have demonstrated translocation of DNA molecules through graphene nanogaps and observed signatures that have features different from those observed with circular solid-state nanopores. We argue that these derive from the DNA interacting with the unique nanogap geometry and we conclude that DNA-graphene gap interactions are important and need to be included in a realistic design of graphene-nanogap based sequencing devices. This study opens up new advancement in single molecule genomic screening devices, and DNA sequencing.

\section*{Acknowledgments}
We acknowledge  support from the National Science Foundation under award DMR-1034937. We thank Richard Kaner and Jaime Torres for discussions. We thank Cees Dekker and Greg Schneider for useful discussions and their hospitality while helping us learn their wedge-transfer technique. 

\nolinenumbers

% Either type in your references using
% \begin{thebibliography}{}
% Text
% \end{thebibliography}
%
% or
%
% Compile your BiBTeX database using our plos2015.bst
% style file and paste the contents of your .bbl file
% here.
% 

\bibstyle{plos2015}
\bibliography{references}

\section*{Supporting information}

{\bfseries S1 File. Source data.} This is the data described in the manuscript.

\end{document}